# All-optical atomic magnetometers based on nonlinear magneto-optical rotation with amplitude modulated light

Szymon Pustelny[a], Adam Wojciechowski[a], Mateusz Kotyrba[a], Krystian Sycz[a], Jerzy Zachorowski[a], Wojciech Gawlik[a],

[a] Centrum Badań Magnetooptycznych, M. Smoluchowski Institute of Physics, Jagiellonian University, 30-059 Kraków, Poland;

Arman Cingoz[b], Nathan Leefer[b], James M. Higbie[b], Eric Corsini[b], Micah P. Ledbetter[b], Simon M. Rochester[b], Alexander O. Sushkov[b], and Dmitry Budker[b]

[b] Department of Physics, University of California at Berkeley, Berkeley, CA 94720-7300, USA

## ABSTRACT

We demonstrate a magnetometric technique based on nonlinear magneto-optical rotation using amplitude modulated light. The magnetometers can be operated in either open-loop (typical nonlinear magneto-optical rotation with amplitude-modulated light) or closed-loop (self-oscillating) modes. The latter mode is particularly well suited for conditions where the magnetic field is changing by large amounts over a relatively short timescale.

## 1. INTRODUCTION

### 1.1. The Krakow-Berkeley Laboratory

In this contribution, we present some recent results of the collaboration between our two experimental atomic physics groups at the Jagiellonian University, Kraków and the Department of Physics at the University of California at Berkeley, respectively. For the last several years, the two groups have been involved in scientific exchange including mutual research visits at all levels from undergraduate students to principal investigators, with a distinct emphasis on the student training and providing the students with a unique experience of performing research in a foreign country as a part of an international research team. A highlight of the collaboration is the establishment of a joint Kraków-Berkeley laboratory (www.if.uj.edu.pl/pl/ZF/k-b_lab/) located at the M. Smoluchowski Institute of Physics at Kraków, where Berkeley students work on a regular basis in close collaboration with Polish colleagues.

### 1.2. Nonlinear magneto-optical rotation (NMOR)

The central subject of the collaboration is the investigation of nonlinear magneto-optical processes (see, for example, review [1]), including their fundamental properties, as well as applications to sensitive electromagnetic measurements, and, in particular, to atomic magnetometry.

In general, a nonlinear magneto-optical interaction can be thought of as consisting of three major steps. In the first step, light interacting with an atomic ensemble creates anisotropy (polarization) of the ground atomic state, which, in the second step, evolves in the presence of external electromagnetic fields. In most NMOR experiments and in the present work, the dominant contribution to the ground state polarization is the rank 2 polarization (quadrupole) moment, also known as alignment. In the case of a weak static magnetic field and low light power, the evolution is simply the Larmor spin precession of the polarized atoms. In general, the evolution could be more complicated, and could be influenced, for example, by off-resonant interactions with the strong laser light itself, leading to processes such as alignment-to-orientation conversion (AOC) (see, for example, [2]). Finally, the third stage of a magneto-optical rotation involves detecting the evolved atomic polarization with probe light. Either optical absorption or rotation can be used in this last step; however, optical rotation is typically preferred as it allows for cancellation of common-mode optical noise.

In typical Faraday geometry (where the magnetic field is parallel to the propagation direction of linearly polarized light) with unmodulated light, the optical rotation measured as a function of the magnetic field takes the shape of a dispersive Lorentzian centered about zero magnetic field. The small-field magnetometric sensitivity is then inversely proportional to the slope of the optical rotation as a function of the magnetic field, given by $A/\gamma$, where $A$ is the amplitude of the zero field resonance and $\gamma$ is the width of the resonance proportional to the ground state polarization relaxation rate. Sensitivity drops rapidly when the rate of Larmor precession is greater than the ground state relaxation rate because the light-induced polarization gets washed out due rapid precession in the magnetic field.

To extend the magnetometric sensitivity to magnetic fields where Larmor precession is much faster than the ground state relaxation rate, it is necessary to synchronize the optical pumping rate with Larmor precession. This can be achieved by modulating the light (see, for example, review [3]). The two most commonly used types of modulation are those of frequency (FM) and amplitude (AM). Both these methods allow one to construct sensitive all-optical, broad dynamic range devices that do not require application of any additional fields to the atoms (such as the radio- or microwave-frequency magnetic fields used in traditional double-resonance optical-pumping magnetometers). The latter method is the subject of this contribution.

Pioneering applications of amplitude-modulated light were introduced in atomic and molecular spectroscopy in the early 1960's by Bell and Bloom [4] and Corney and Series [5]. In the context of NMOR it was done in Refs. [6,7]. The application of amplitude-modulated light in NMOR, the so-called AMOR (amplitude-modulated optical rotation), yields narrow NMOR resonances centered at zero magnetic field, along with additional sets of resonances at higher magnetic fields where the frequency of amplitude modulation is synchronized with the Larmor precession. The position of these high-field resonances is linear in the frequency of amplitude modulation $\Omega_m$ and their amplitude and width is comparable with the zero-field resonances, yielding the possibility of precise measurements of strong magnetic fields.

## 2. EXPERIMENTAL SETUP

The layout of AMOR experimental setup is presented in Fig. 1. Light from an external-cavity diode laser was tuned to the center of the $F$=2→$F'$=1 D1 line of $^{87}$Rb (795 nm). Light frequency was controlled with a Doppler-free dichroic lock [8], which enables its precise, long-term stabilization. Some light was also directed to a saturation spectroscopy system, providing information about the frequency of the laser light. The main beam was transmitted in a single pass through an analog acousto-optical modulator (AOM) driven by an 80-MHz radio-frequency (RF) signal. The amplitude of the RF signal could be modulated with arbitrary waveforms and frequencies ranging from ~1 kHz to ~80 kHz producing a corresponding amplitude modulation of the laser light. The source of amplitude modulation was either a function generator in the open-loop mode (see Sec. 3.1), or a signal derived from the measured optical rotation in the self-oscillating mode (see Sec. 3.2).

The amplitude-modulated light of 2 mm in diameter passed through a high-quality linear polarizer and then through a cylindrical, paraffin-coated glass cell (1 cm long, 1 cm in diameter) containing isotopically enriched $^{87}$Rb (nuclear spin $I$=3/2). The cell was heated to about 40°C to achieve optimal concentration of rubidium atoms (the absorption depth of about 0.3). The paraffin coating on the cell greatly reduces the relaxation of ground-state atomic polarization during collisions with the cell walls, yielding lifetimes on the order of hundreds of milliseconds [9]. Since the atoms bounce off the cell walls many times in one polarization-relaxation time, they effectively sample the average magnetic field in the cell, reducing relaxation due to field inhomogeneities [10], an effect known as motional narrowing. The cell was placed within a 3-layer $\mu$-metal magnetic shield. The shield provided passive attenuation of external magnetic fields at a level of about one part in ten thousand. Additionally, a set of three mutually orthogonal magnetic coils was placed inside the shield. It enabled compensation of residual magnetic fields inside the shield, as a well as generation of a well controlled magnetic field. Using the coils the longitudinal magnetic field could be scanned from -200 mG to 200 mG.

After passing through the cell, the polarization of the light was analyzed with a balanced polarimeter that consisted of a polarizing beam-splitter rotated by 45° with respect to the initial light polarization and two photodiodes. In open-loop mode, a photodiode difference signal was fed to a lock-in amplifier and measured at the first harmonic of light-modulation frequency $\Omega_m$. In-phase and quadrature signals were stored on a computer. In the self-oscillating mode with closed-loop, the photodiode difference signal was first amplified and phase-shifted and then fed back to the AOM. The frequency of the resulting signal in this case was measured with a frequency counter.

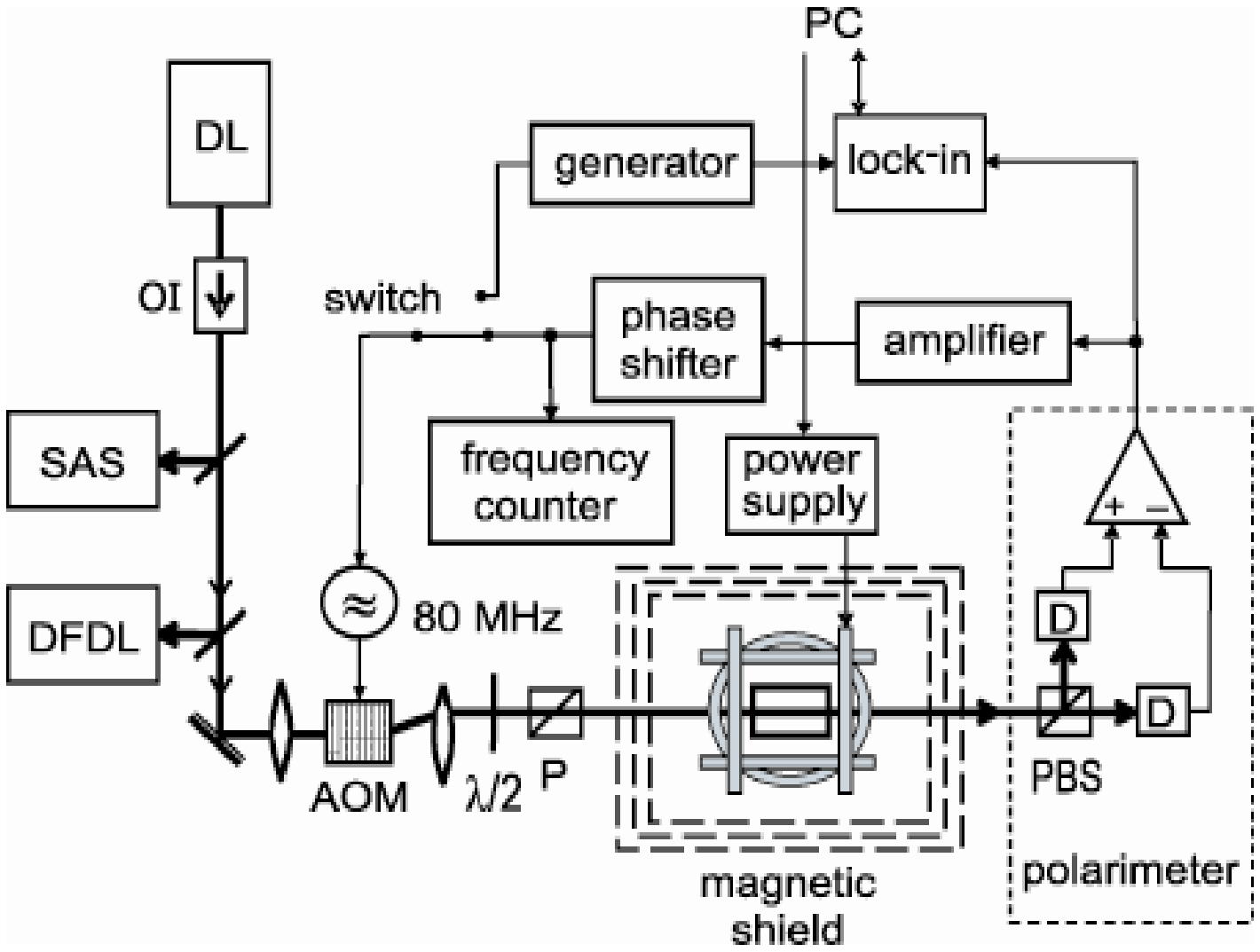

**Figure 1.** AMOR experimental setup. DL – external-cavity diode laser, OI – optical isolator, SAS – saturated absorption system, DFDL – Doppler-free dichroic lock, λ/2 – half-wave plate, D – photodiodes, P – crystal polarizer, PBS – polarization beam-splitter, PC – computer controlling the experiment. With a switch in one position the system worked as AMOR in self-oscillating mode with closed loop. Changing the switch to the other position converted the experiment to the open-loop AMOR experiment (open-loop configuration.

## 3. RESULTS

### 3.1. Open-loop AMOR configuration

Typical non-optimized AMOR signals with the open-loop configuration are presented in Fig. 2. The results were recorded with sinusoidal modulation of 100% depth and with an average light intensity, e.g., intensity measure after AOM, of 3 μW/mm$^2$. In addition to a typical zero field resonance observed in unmodulated NMOR experiments, two resonances were observed centered at magnetic fields $B$ such that

$$\Omega_m = \pm 2\Omega_L = \pm 2g\mu_B B/\hbar, \tag{1}$$

where $\Omega_L$ is the Larmor frequency, $g \approx 2/(2I+1)$ is the Landé factor, $\mu_B$ is the Bohr magneton. Given the linear relationship between the center of the resonance and the magnetic field, the AMOR resonance can be used to directly obtain a measurement of the magnetic field. Using frequency-modulated light yields resonances in the same position [11] and it was recently demonstrated that sensitivity on the order of $6\cdot10^{-10}$ G/Hz$^{1/2}$ in a 3.5-cm cell can be obtained in geophysical range fields in open-loop mode [12].

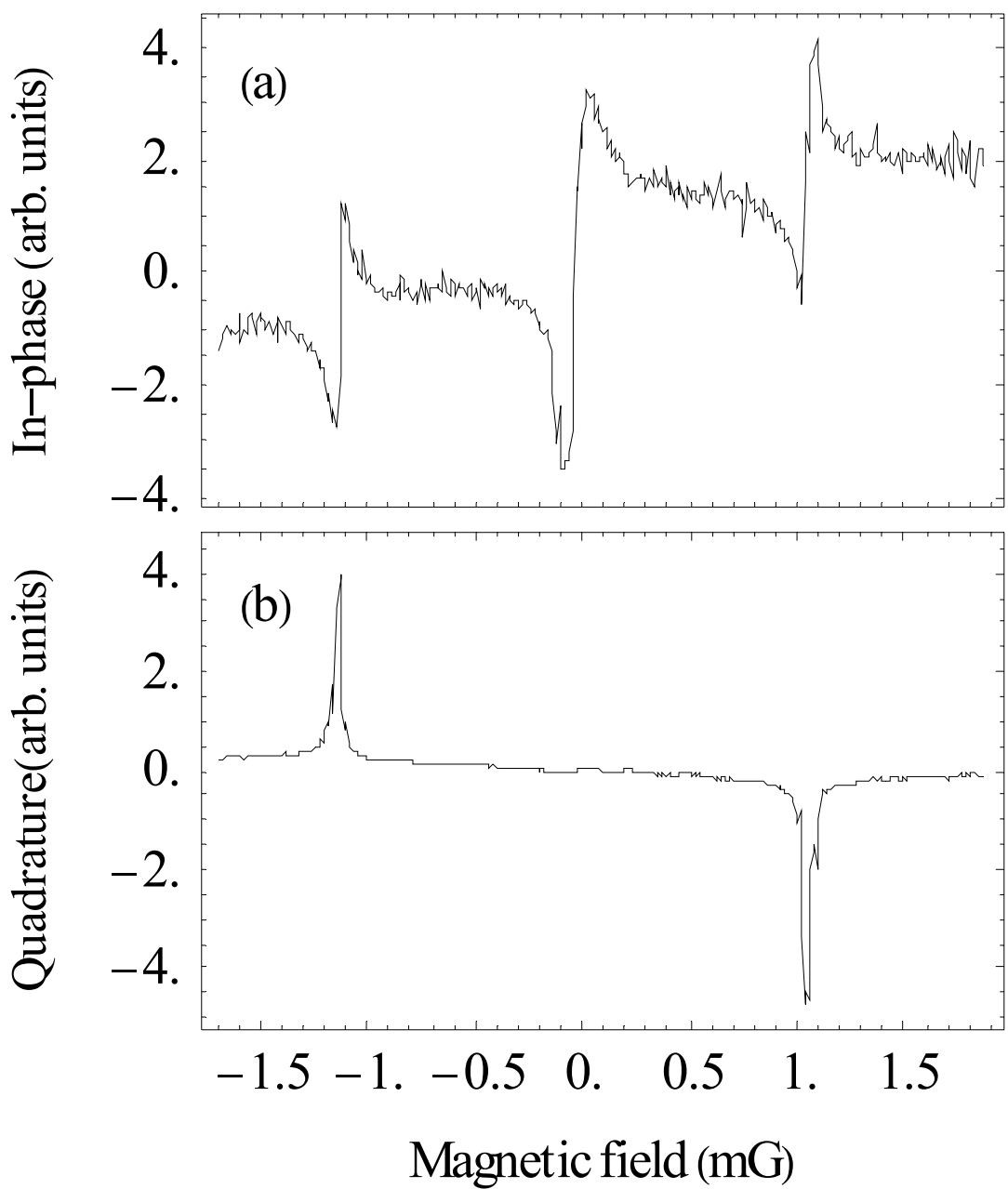

**Figure 2.** Synchronously detected in-phase (a) and quadrature (b) AMOR signals. Light modulation was sinusoidal with modulation depth of 100% and the average light intensity was 3 μW/mm$^2$.

One advantage of AM over FM is the straightforward possibility of varying the modulation waveforms. In particular, AM enables modulating light intensity with variable duty cycle square-modulation. In Fig. 3 we show the AMOR signals for several different duty cycles of square-wave modulation. Decreasing the duty cycle caused appearance of additional resonances at magnetic fields corresponding to $\Omega_L = n\Omega_m/2$, where $n$ is an integer number. These resonances arise due to higher harmonic of $\Omega_m$ present in the modulation signal. While for 50% duty cycle the dominant Fourier component in the modulation signal is the one at $\Omega_m$, for lower duty cycles the higher Fourier components ($n\Omega_m$) become significant, leading to strong additional resonances. To verify that our understanding of this phenomenon is correct we simulated signals using a modified, time-dependent Kanorsky-Weis model [13]. A dashed line in Fig. 3(c) shows predictions of the model for a given duty cycle. The predictions are in good agreement with experimental results.

In a discussed case resonances occur at harmonics of the modulation frequency and they reach high magnetic fields even with low modulation frequency of light. This reduces experimental requirements because high magnetic fields can be measured with low frequency and low duty cycle. It also enables the stabilization of the magnetic fields so that the Larmor frequency and the modulation frequencies are simple fractions, for example, 2:1, 3:5, 7:2.

An advantage of AMOR over FM NMOR is a possibility of tuning laser frequency in order to minimize AC Stark effect [6]. This effect is one of the limiting factors in high-precision measurements of magnetic fields based on NMOR.

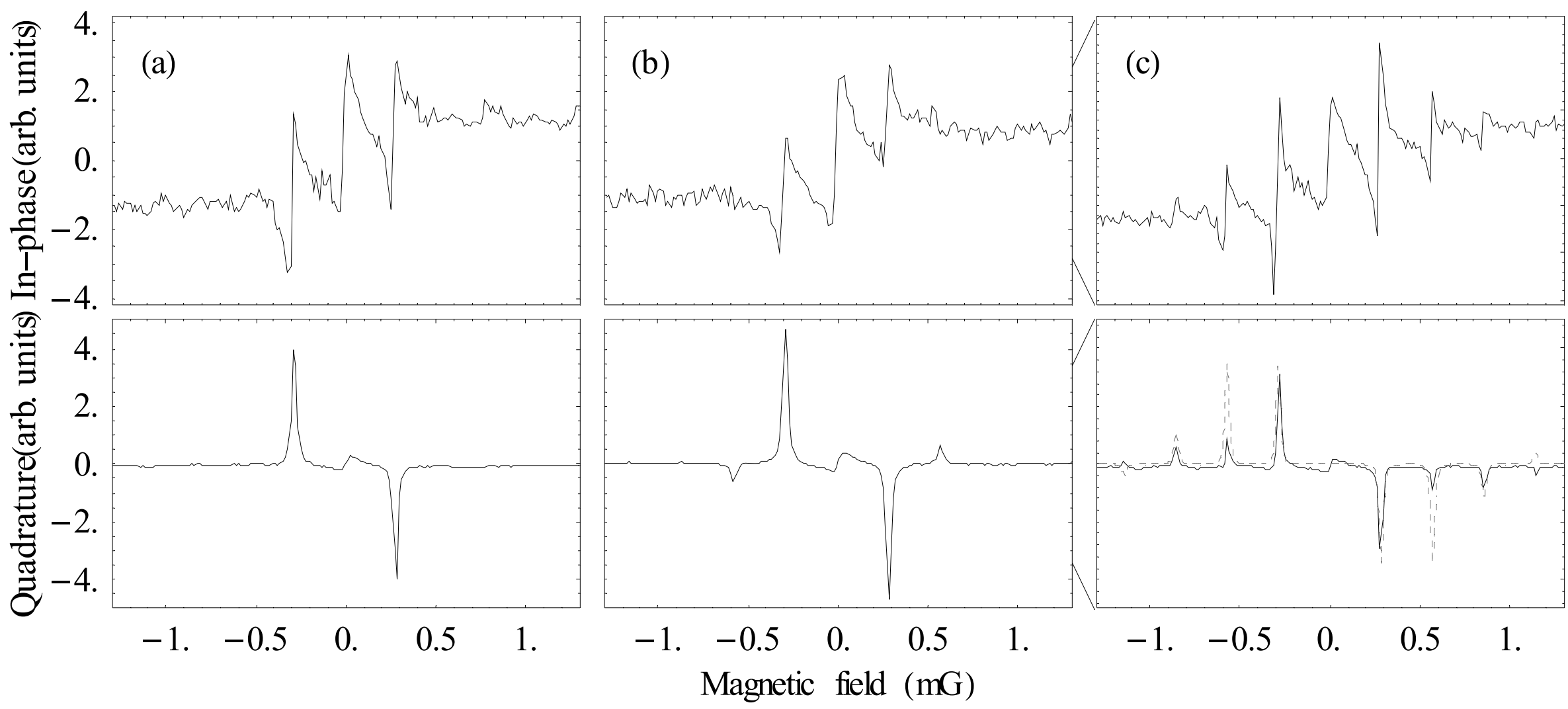

**Figure 3.** Synchronously detect AMOR signals recorded with 100% square-wave modulation and 0.5 duty cycle and 15 μW average power (a), 0.35 duty cycle and 10 μW average light power (b), and 0.2 duty cycle and 6 μW average light power (c). Upper row shows in-phase signals, lower row shows quadrature signals. The dashed line in quadrature signal of 0.2 duty cycle shows predictions of the model. Small discrepancy between experimental curve and theoretical predictions shown in (c) is due to not perfectly square modulation of light.

## 3.2. Self-oscillation AMOR (closed-loop configuration)

In a typical AMOR experiment, the optical rotation is detected synchronously with a lock-in amplifier at the first harmonic of $\Omega_m$. If the magnetic field varies by less than the width of the resonance, the resulting in-phase part of the signal can give a direct measurement of the magnetic field without changing $\Omega_m$. However, since very narrow resonances can be obtained with paraffin coated cells, this results in a magnetometer with a very small dynamic range centered about $B$=0 and $B=\pm\hbar\Omega_m/(2g\mu_B)$. For a larger dynamic range, the in-phase component may be used as an error signal, adjusting the modulation frequency so that the in-phase component is zero. The magnetic field can then be determined from Eq. (1). The major drawback of this approach is that if the magnetic field jumps by many linewidths, the resonance will be lost.

A more practical approach is the so-called self-oscillation mode, originating in the classic work [14]. In our version of this approach, instead of using an external function generator to drive the AOM, the difference signal from the balanced polarimeter drives the AOM directly. We start by assuming some initially aligned atomic state and some weak, static laser beam. The atoms precess in the magnetic field, generating optical rotation and photocurrent at $2\Omega_L$. This photocurrent drives the AOM, modulating the light at the exact frequency necessary to reinforce the initial alignment. In this manner, the entire system which includes the polarized atoms, polarimeter, and amplitude modulated light becomes an oscillator, oscillating at $2\Omega_L$. The onset of self-oscillation can be analyzed by considering some initial noise in the laser or balanced polarimeter to jump-start the oscillator.

The above scheme can be implemented with either a single beam, as in the present work and shown in Fig. 1, or with two beams, as in Ref. [15]. In either case, maximal optical rotation of the probe beam occurs when the aligned state is rotated by $\pi/4$ with respect to the initial polarization of the light. Since alignment is created parallel to the initial polarization of the light, it is thus necessary to shift the phase of the signal from the balanced polarimeter by $\pi/2$ before feeding it back into the AOM. In the single-beam arrangement, this is accomplished electronically with the phase shifter in Fig. 1, while in the two-beam configuration this is accomplished mechanically by rotating the polarization of the probe beam by $\pi/4$ relative to the pump beam [15].

In Fig. 4 we show an example of single-beam self-oscillating AMOR signal for typical parameters. The dominant Fourier component of the signal is at $2\Omega_L$, with additional higher harmonics. The primary reason for the presence of higher harmonics is the fact that the functions of pumping and probing are accomplished by the same beam. Hence, the amplitude of the "probe" beam is also being modulated periodically (the dominant Fourier component is at $2\Omega_L$, but higher harmonics are also present) in addition to the rotation of the polarization. There may also be some effects from the electronics in the feedback loop. In future work these issues will be addressed in more detail.

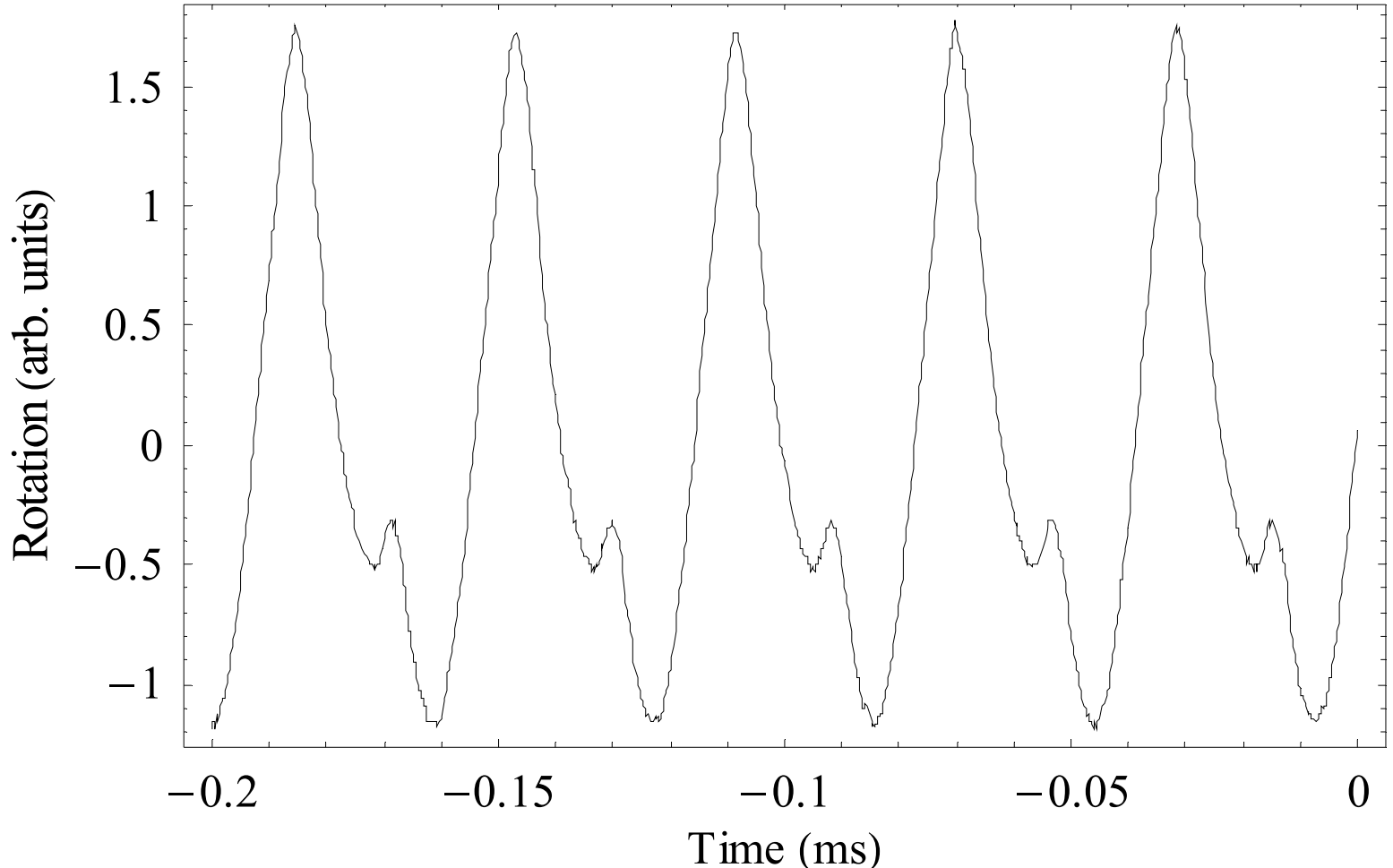

**Figure 4.** A typical single-beam self-oscillation AMOR signal. The oscillations were recorded for relatively high light intensity ~35 μW/mm$^2$.

Despite the complex structure of the self-oscillating AMOR signal, its frequency can be still measured. In Fig. 5 the dependence of the AMOR signal frequency versus magnetic field is shown. The frequency was measured directly with a frequency counter without any filtering. The frequency of the self oscillating AMOR signal is linear in the magnetic field, as expected.

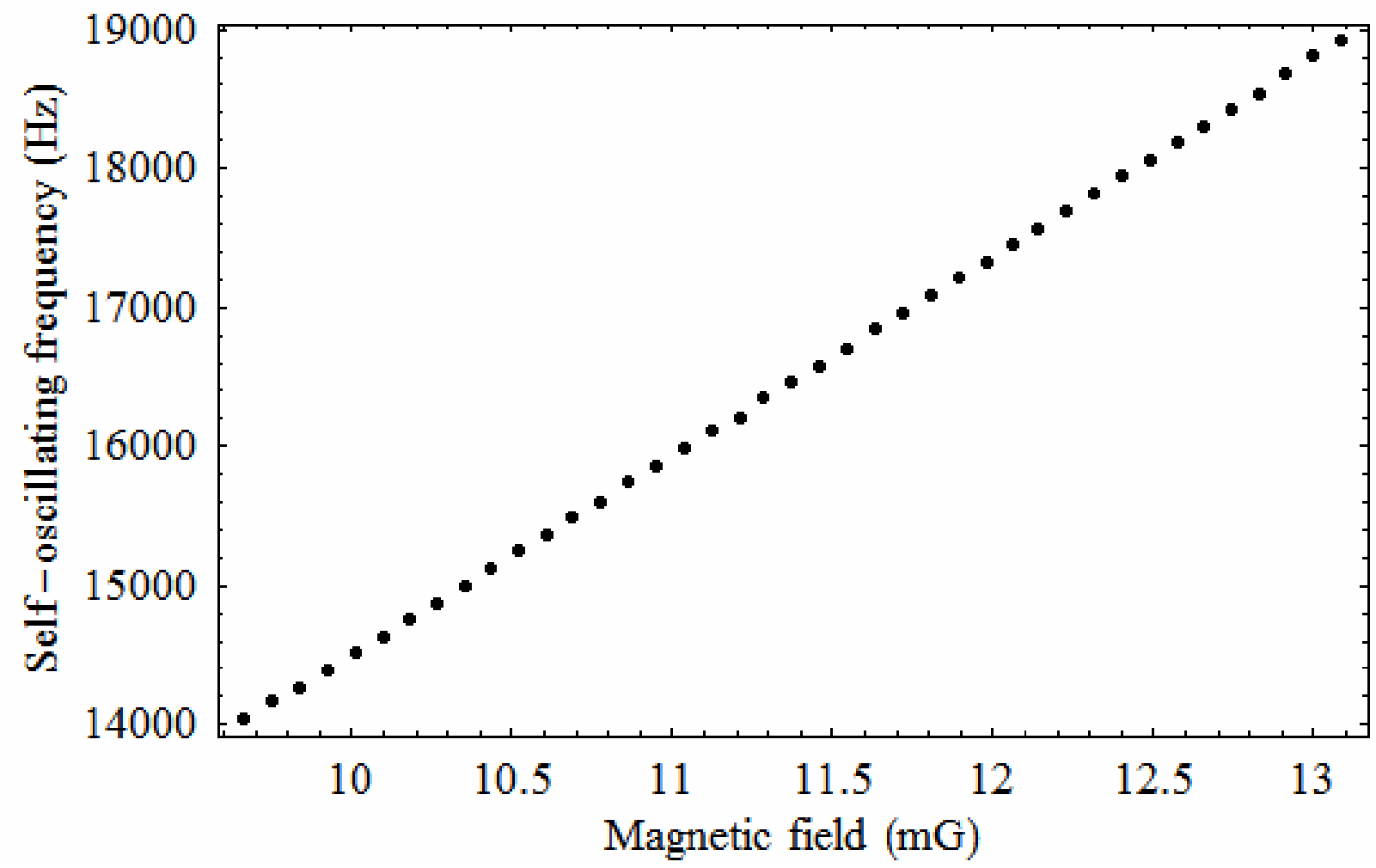

**Figure 5.** Frequency of the single-beam self-oscillation AMOR signal follows changes of the magnetic field. Results were recorded with light intensity of ~35 μW/mm$^2$.

The range of magnetic fields over which the present experiment can operate is relatively small, limited by the electronics used to shift the phase of the signal before being fed back into the AOM. To achieve sensitivity to a larger range of magnetic fields it will be necessary to use a phase shifter with a wider bandwidth. An alternative solution may be the use of a phase-locked loop as in recent work of Ref. [16] on a self oscillating magnetometer with frequency-modulated light. Such a solution would ensure the constant phase-shift between light modulation and AMOR signal independently of the frequency of these signals. As mentioned above, the needed phase shift is easily produced the in a two-beam arrangement by rotating the polarization of the pump beam relative to the probe. In addition to producing a frequency independent phase shift, this has the added advantage that the resulting signal from the balanced polarimeter is purely sinusoidal.

One of the advantages of the self-oscillating magnetometers is its bandwidth. The demonstrated bandwidth of the self-oscillating two-beam magnetometer exceeds 1 kHz [15]. In addition with a sensitivity at a level of 3 nG achieved for 1-s measurement [15], it opens a possibility of construction of a high-performance all-optical atomic magnetometer. Application of such components as vertical-cavity surface-emitting lasers, optical fibers, microfabricated vapor cells, ect., enables miniaturization of the device and its low-power consumption. Thus it opens a possibility for its commercial applications.

Figure 6 shows quasi-instantaneous response of the self-oscillating magnetometer to a change of the magnetic field. The response of the two-beam self-oscillating magnetometer was monitored by heterodyning the NMOR signal with a fixed reference frequency on a lock-in amplifier (for more details see Ref. [15]).

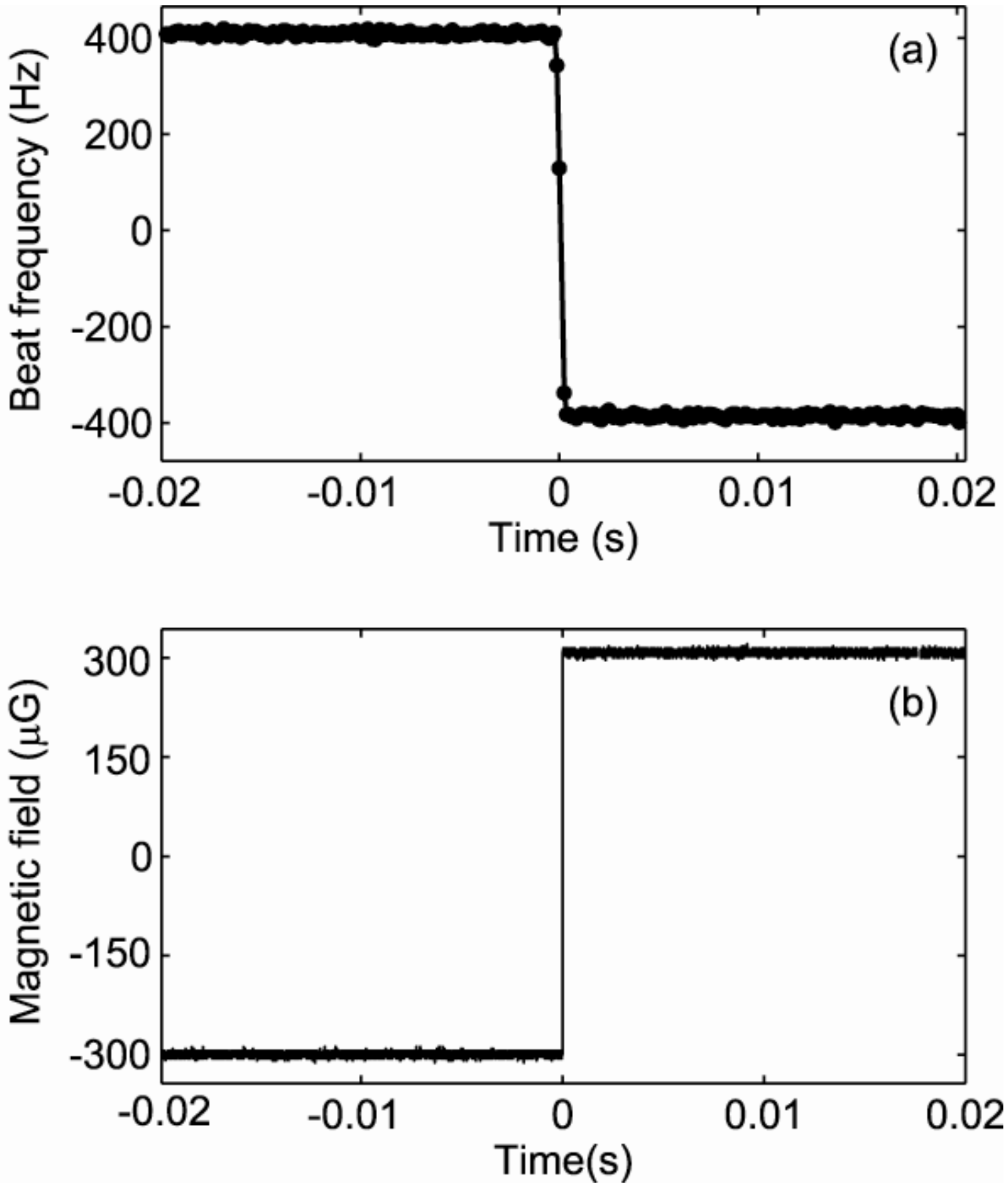

Fig. 6. Response of the two-beam self-oscillating magnetometer to a 570 μG step in magnetic field, in the presence of an over-all bias field of ~14mG (a). The step is large compared to the resonance line width, but small compared to the bias field. The self-oscillation waveform was recorded on a digital storage oscilloscope and subsequently fit to a sinusoid in overlapping time windows 500 μs long, spaced by 125 μs. In part (b), step change of a magnetic field is shown.

## 4. CONCLUSION

We have demonstrated operation of all-optical atomic magnetometers based on nonlinear magneto-optical rotation with amplitude-modulated light. First a typical AMOR experiment was discussed and its possible applications for magnetic field measurements were described. Then attention was switched to AMOR experiments in the so-called self-oscillation regime. In such configuration NMOR signal is used for amplitude modulation of light which simplifies an experimental arrangement setup. Two different experimental realizations of self-oscillation AMOR were presented. First scheme uses a single light beam for pumping and probing atoms. That facilitates the use of microfabrication techniques to construct small, lightweight, portable atomic magnetometers. The range of magnetic fields over which the present experiment may self oscillate is limited by the bandwidth of the electronics used to produce the necessary phase shift. Finally, a robust, high-bandwidth magnetometer using separated (modulated) pump and (unmodulated) probe laser beams has been described. In such arrangement magnetic field can be measured at a sensitivity level 3 nG at 1 s with a bandwidth 1 kHz.

## ACKNOWLEDGMENTS

The authors acknowledge invaluable contributions of E. B. Alexandrov, M. V. Balabas, D. F. Jackson Kimball, C. J. Hovde, S. Xu, and V. V. Yashchuk. The work presented here was supported in part by US DOD MURI grant \# N-00014-05-1-0406, KBN grant \# 1 P03B 102 30, an ONR STTR grant through Southwest Sciences, Inc.. Additionally, the Berkeley-Krakow collaboration has been supported by NSF US-Poland collaboration grant, and NATO Linkage Grant. One of the authors (S.P.) is a scholar of the project co-financed from the European Social Fund.

## REFERENCES

1. D. Budker, W. Gawlik, D.F. Kimball, S.M. Rochester, V.V. Yashchuk, A. Weis, "Resonant nonlinear magneto-optical effects in atoms", *Rev. Mod. Phys.* **74**, pp. 1153-1201, 2002.
2. D. Budker, D. F. Kimball, S. M. Rochester, and V. V. Yashchuk, "Nonlinear Magneto-optical Rotation via Alignment-to-Orientation Conversion", *Phys. Rev. Lett.* **85**, pp. 2088-2091, 2000.
3. E. B. Alexandrov, M. Auzinsh, D. Budker, D. F. Kimball, S. M. Rochester, and V. V. Yashchuk, "Dynamic effects in nonlinear magneto-optics of atoms and molecules", *JOSA B* **22**, pp. 7-20, 2005.
4. W. E. Bell and A. L. Bloom, "Optically driven spin precession", *Phys. Rev. Lett.* **6**, pp. 280-281, 1961.
5. A. Corney and G. W. Series, "Theory of resonance fluorescence excited by modulated or pulsed light", *Proc. Phys. Soc.* **83**, pp. 207-212, 1964.
6. W. Gawlik, L. Krzemień, S. Pustelny, D. Sangla, J. Zachorowski, M. Graf, A. O. Sushkov, and D. Budker, "Nonlinear magneto-optical rotation with amplitude modulated light", *Appl. Phys. Lett.* **88**, pp. 1311081-13110813, 2006.
7. M. V. Balabas, D. Budker, J.Kitching, P.D.D. Schwindt and J. E. Stalnaker, "Magnetometry with millimeter-scale antirelaxation-coated alkali-metal vapor cells", JOSA B **23**, pp. 1001-1006, 2006.
8. G. Wąsik, W. Gawlik, J. Zachorowski, and W. Zawadzki, "Laser frequency stabilization by Doppler-free magnetic dichroism", *Applied Physics B* **75**, pp. 613-619 (2002).
9. D. Budker, L. Hollberg, D. F. Kimball, J. Kitching, S. Pustelny, and V. V. Yashchuk, "Investigation of microwave transitions and nonlinear magneto-optical rotation in anti-relaxation-coated cells", Phys. Rev. A **71**, pp. 0120931-0120939 (2005).
10. S. Pustelny, D. F. Jackson Kimball, S. M. Rochester, V. V. Yashchuk, and D. Budker, "Influence of magnetic-field inhomogeneity on nonlinear magneto-optical resonances" – accepted by Phys. Rev. A.
11. D. Budker, D.F. Kimball, V.V. Yashchuk, and M. Zolotorev, "Nonlinear magneto-optical rotation with frequency-modulated light", Phys. Rev. A **65**, 055403, 2002.
12. V. Acosta, M. P. Ledbetter, S. M. Rochester, D. Budker, D. F. Jackson-Kimball, D. C. Hovde, W. Gawlik, S. Pustelny, and J. Zachorowski, V. V. Yashchuk, "Nonlinear magneto-optical rotation with frequency-modulated light in the geophysical field range", *Phys. Rev. A* **73**, pp. 0534041-0534048, 2006.
13. S. I. Kanorsky, A. Weis, J. Wurster, and T. W. Hänsch, "Quantitative investigation of the resonant nonlinear Faraday effect under conditions of optical hyperfine pumping", *Phys. Rev. A* **47**, pp. 1220-1226, 1993.
14. A. L. Bloom, *Appl. Opt.* **1**, pp. 61-68, 1962.
15. J. M. Higbie, E. Corsini, and D. Budker, "Robust, High-speed, All-optical Atomic Magnetometer" - submitted to *Rev. Sci. Instr.*
16. P. D. D. Schwindt, L. Hollberg, and J. Kitching, *Rev. Sci. Instr.* **76**, pp. 1261031-1261034, 2005.